\documentclass[twocolumn,prl,superscriptaddress]{revtex4}
\usepackage{amsmath}
\usepackage{amssymb}
\usepackage{graphicx}
\usepackage{esint}
\usepackage{pdfpages}

\makeatletter
\@ifundefined{textcolor}{}
{%
 \definecolor{BLACK}{gray}{0}
 \definecolor{WHITE}{gray}{1}
 \definecolor{RED}{rgb}{1,0,0}
 \definecolor{GREEN}{rgb}{0,1,0}
 \definecolor{BLUE}{rgb}{0,0,1}
 \definecolor{CYAN}{cmyk}{1,0,0,0}
 \definecolor{MAGENTA}{cmyk}{0,1,0,0}
 \definecolor{YELLOW}{cmyk}{0,0,1,0}
}

\newcommand{\ignore}[1]{}

\@ifundefined{textcolor}{}{%
 \definecolor{BLACK}{gray}{0}
 \definecolor{WHITE}{gray}{1}
 \definecolor{RED}{rgb}{1,0,0}
 \definecolor{GREEN}{rgb}{0,1,0}
 \definecolor{BLUE}{rgb}{0,0,1}
 \definecolor{CYAN}{cmyk}{1,0,0,0}
 \definecolor{MAGENTA}{cmyk}{0,1,0,0}
 \definecolor{YELLOW}{cmyk}{0,0,1,0}
}

\newcommand{\rmc}{{\rm c}}
\newcommand{\rmd}{{\rm d}}
\newcommand{\rme}{{\rm e}}

\newcommand{\rmi}{{\rm i}}

\makeatother

\begin{document}

\title{Band-edge superconductivity}

\author{Garry Goldstein}

\affiliation{Department of Physics, Rutgers University, Piscataway, New Jersey
08854, USA}

\author{Camille Aron}

\affiliation{Department of Electrical Engineering, Princeton University, Princeton,
New Jersey 08544, USA}

\author{Claudio Chamon}

\affiliation{Department of Physics, Boston University, Boston, Massachusetts 02215,
USA}

\begin{abstract}
We show that superconductivity can arise in semiconductors with a
band in the shape of a Mexican hat when the chemical potential is
tuned close to the band edge, but not intersecting the band, as long
as interactions are sufficiently strong. Hence, this is an example
where superconductivity can emerge from a band insulator when interactions
exceed a threshold. Semiconductors with simple cubic symmetry point
groups and with strong spin-orbit coupling provide an example of a
system with such band dispersion. 
\end{abstract}

\maketitle

The BCS theory of superconductivity is perhaps the most successful
mean-field theory~\cite{BCS1,BCS2}. It explains the phenomenology
of many known superconductors, although, notably, it fails to describe
the cuprate high $T_{c}$ superconductors. BCS theory takes as starting
point a good metal, with a sizable Fermi sea, and then explains the
formation of the Cooper pairs at the Fermi surface, mediated by the
electron-phonon interaction. Because the Cooper pairs occur only on
a thin momentum shell, one may wonder if there may be a more ``economical''
way to form the pairs, without the sizable filled Fermi sea.

\begin{figure}
\begin{centering}
\includegraphics[scale=0.6]{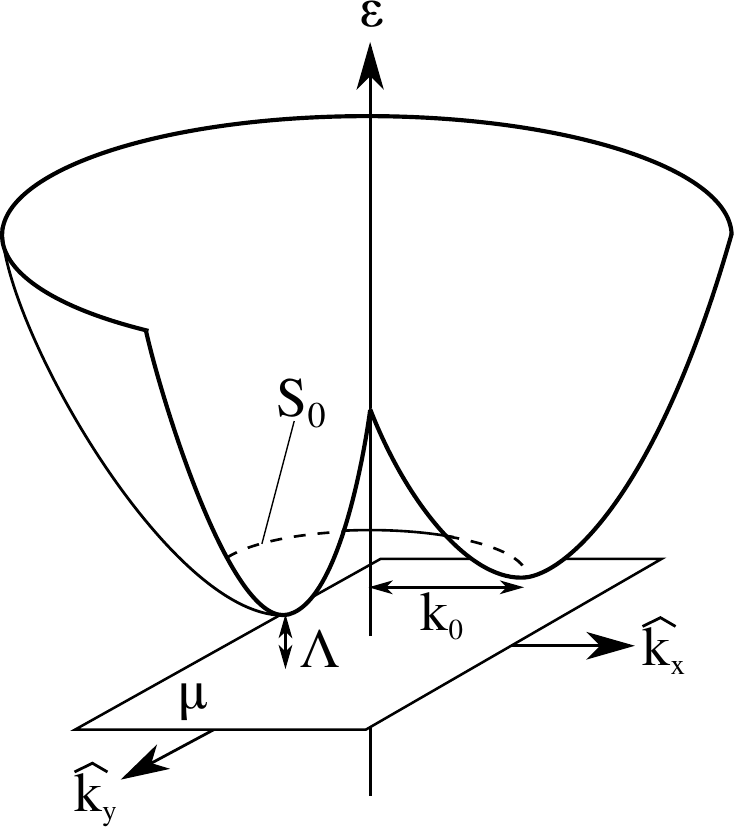} 
\par\end{centering}

\protect\protect\protect\protect\protect\protect\caption{\label{fig:extremum} Semiconductor with a rotationally symmetric
band extremum (here minimum). The chemical potential is tuned right
below the band edge, at distance $\Lambda$ from the bottom.}
\end{figure}

In this paper, we start with a system \textit{without} a Fermi surface,
a semiconductor where the chemical potential does not intercept the
dispersing band. In the absence of interactions, this system has zero
conductivity at zero temperature. We show, however, that for certain
geometries of bands near the band edge, interactions can lead to superconductivity.
The favorable dispersion (band geometry) is when the locus of the
band edge in the Brillouin zone is not a single point (as in a parabolic
band), but instead is a $d-1$ dimensional momentum \emph{shell} ${\cal S}_{0}$
(in the case of a $d$-dimensional semiconductor). The electrons in
this shell are those responsible for superconductivity in the presence
of interactions. Fig.~(\ref{fig:extremum}) depicts the relevant
situation, showing a Mexican hat dispersion and the chemical potential
just missing the edge of the band. Near the extrema, the density of
electronic states scales as in a one-dimensional system as long as
the radius of the momentum shell ${\cal S}_{0}$ extremum is non-zero.
As we shall see, the interactions (when sufficiently large) are responsible
for a non-trivial occupation of the shell even when the chemical potential
does not cross the bands. Hence, at non-zero temperature this system
undergoes a superconducting phase transition as a function of the
interaction strength. The superconducting transition temperature depends
on the strength of the interactions and on the detuning of the chemical
potential from the band edge. We find that such systems can have rather
large transition temperatures, possibly on the order of room temperature
for reasonable interaction strengths.

In~\cite{Garry1}, while discussing the possibility of inducing superconductivity
in large-gap semiconductors by shining an ac electric field, we had
emphasized the importance of the Mexican hat geometry of the effective
band which emerges in the rotating frame. Such a geometry also arises
in driven Dirac systems~\cite{Driven-masses}. But this geometry
is not tied to a driven system and it can also be found in equilibrium
situations such as non-centrosymmetric systems with spin-orbit interaction
and cubic symmetry, such as Li$_{2}$(Pd$_{1-x}$Pt$_{x}$)$_{3}$B~\cite{key-1,key-14,he-xi-lu}.
In the latter case, for instance, the single particle Hamiltonian
reads (we set $\hbar=1$) 
\begin{equation}
{\cal H}_{\boldsymbol{k}}=\frac{{k}^{2}}{2m^{*}}+\alpha\,\boldsymbol{k}\cdot\boldsymbol{\sigma}\;,\label{eq:H-non-centrosymmetric}
\end{equation}
where $k\equiv|\boldsymbol{k}|$, $\alpha$ is the spin-orbit coupling,
$\boldsymbol{\sigma}=(\sigma_{1},\sigma_{2},\sigma_{3})$ are the
Pauli matrices acting on the spin degrees of freedom, and $m^{*}$
is an effective band mass (positive or negative, depending on the
band). The dispersion relation $\epsilon(\boldsymbol{k})=\frac{1}{2m^{*}}k^{2}\pm\alpha k$
has extrema located on $\mathcal{S}_{0}$, the $d-1$-sphere of radius
\begin{equation}
k_{0}=|m^{*}|\,|\alpha|\;,\label{eq:k_0}
\end{equation}
see Fig.~(\ref{fig:extremum}). The near-extrema dispersion relation
is quadratic, $\epsilon(\boldsymbol{k})\approx\frac{1}{2m^{*}}(k-k_{0})^{2}+\epsilon_{0}$,
resulting in a density of states which diverges in the same fashion
as in a 1D system, as noted in Ref.~\cite{Driven-masses,key-1,he-xi-lu},
even for 2D or 3D crystals: $\rho_{{\rm 3D}}(\epsilon)=4\pi k_{0}^{2}\,\sqrt{2|m^{*}|}/\sqrt{|\epsilon-\epsilon_{0}|}$
or $\rho_{{\rm 2D}}(\epsilon)=2\pi k_{0}\,\sqrt{2|m^{*}|}/\sqrt{|\epsilon-\epsilon_{0}|}$.
If the Fermi energy is set near $\epsilon_{0}$ so as to just cross
the bands slightly, the Fermi surface consists of two concentric spherical
shells (in 3D) or two concentric circles (in 2D) at momenta $k_{0}\pm\delta k_{F}$.
However, we shall concentrate instead in the case where the chemical
potential does not cross the band, and therefore there is no Fermi
surface and consequently either no occupation or complete occupation
of the bands (at zero temperature) in the absence of interactions.
All the phenomena discussed below emerge because of interactions.

The rest of the paper is organized as follows. First, we use a mean-field
analysis to derive the conditions for $p$-wave superconductivity
in the case of systems with large spin-orbit coupling. Then, we address
the case without spin-orbit coupling, which is pertinent to a situation
where the Mexican hat potential may arise by some other mechanism.
In this case, $s$-wave paring is allowed. Finally we go beyond weak-coupling
mean-field theory and compute the critical temperature $T_{\rmc}$
by means of a calculation \textit{à la} Migdal-Eliashberg~\cite{Eliashberg},
\textit{i.e.} by considering the interplay between Coulombic repulsive
interactions and phonon-mediated attractive interactions.

\paragraph{\label{sec:Mean-field-analysis}Mean-field analysis.}

For the sake of generality, let us depart from the Hamiltonian in
Eq.~(\ref{eq:H-non-centrosymmetric}) and simply assume that the
semiconductor has one relevant band with dispersion $\epsilon\left(\boldsymbol{k}\right)$
and a shallow extremum (either a band maximum or band minimum) located
on a surface ${\cal S}_{0}$. For all wave vectors $\boldsymbol{k}_{0}\in{\cal S}_{0}$,
$\epsilon\left(\boldsymbol{k}_{0}\right)\approx\mbox{cte}$, see Fig.~(\ref{fig:extremum}).
Additionally, let us assume that $\epsilon(\boldsymbol{k})=\epsilon\left(-\boldsymbol{k}\right)$.
Note that this does not necessarily require inversion symmetry. The
spin structures at $\boldsymbol{k}$ and $-\boldsymbol{k}$ are locked
to the respective momenta due to the spin-orbit coupling, and hence
we drop any reference to spins, which do not play any role in what
follows. Also, since the cases with a band maximum or a band minimum
are alike, let us consider only the case of a band minimum ($m^{*}>0$)
with the following Hamiltonian 
\begin{align}
H\!=\!\!\int\!(\rmd\boldsymbol{k})\epsilon(\boldsymbol{k})c_{\boldsymbol{k}}^{\dagger}c_{\boldsymbol{k}}+\!\!\int\!(\rmd\boldsymbol{k})(\rmd\boldsymbol{k}')V_{\boldsymbol{k},\boldsymbol{k}'}c_{\boldsymbol{k}}^{\dagger}c_{-\boldsymbol{k}}^{\dagger}c_{\boldsymbol{k}'}^{\,}c_{-\boldsymbol{k}'}^{\,},\label{eq:Main_Hamiltonian}
\end{align}
in which $c_{\boldsymbol{k}}$ is the annihilation operator of an
electron with momentum $\boldsymbol{k}$ and we use the shorthand
notation $(\rmd\boldsymbol{k})\equiv\rmd^{d}\boldsymbol{k}/\left(2\pi\right)^{d}$.
To further simplify the discussion, let us assume that close to ${\cal S}_{0}$
the band structure $\epsilon(\boldsymbol{k})$ and the interaction
term $V_{\boldsymbol{k},\boldsymbol{k}'}$ are isotropic to leading
order so that $\epsilon(\boldsymbol{k})=\epsilon\left(k\right)$ and
$V_{\boldsymbol{k},\boldsymbol{k}'}=V(\hat{\boldsymbol{k}}\cdot\hat{\boldsymbol{k}}')$
can be decomposed into spherical harmonics 
\begin{equation}
V_{\boldsymbol{k},\boldsymbol{k}'}=\sum_{l=0}^{\infty}\left(2l+1\right)V_{l}\;P_{l}(\hat{\boldsymbol{k}}\cdot\hat{\boldsymbol{k}}'),\label{eq:Potential}
\end{equation}
where $\hat{\boldsymbol{k}}\equiv\boldsymbol{k}/k$, $P_{l}$ is the
Legendre polynomial of degree $l$ and $V_{l}\equiv\int_{0}^{1}\rmd\cos\theta\,V\left(\cos\theta\right)P_{l}\left(\cos\theta\right)$.

Since there is only one band only $p$-wave superconductivity is allowed.
For $p$-wave superconductivity, the symmetry of the order parameter,
$\Delta\left(\boldsymbol{k}\right)=-\Delta\left(-\boldsymbol{k}\right)$,
implies that only odd $l$ are appropriate for pairing. The dominant
pairing mechanism is given by the $l=1$ channel and we may approximate
the pairing potential as 
\begin{equation}
V_{\boldsymbol{k},\boldsymbol{k}'}\approx-\frac{3g}{\mathcal{V}}\;\hat{\boldsymbol{k}}\cdot\hat{\boldsymbol{k}}',\label{eq:Simplified potential}
\end{equation}
where $\mathcal{V}$ is the volume of the sample and $g$ is a coupling
constant. For $l=1$ pairing, the most general order parameter is
of the form~\cite{key-11} $\Delta\left(\boldsymbol{k}\right)=\boldsymbol{\Delta}\cdot\hat{\boldsymbol{k}}$
where $\boldsymbol{\Delta}$ is a constant vector. Choosing the order
parameter with the fewest nodes, we consider $\Delta\left(\boldsymbol{k}\right)=\Delta\left(\hat{k}_{x}+\rmi\hat{k}_{y}\right)$
where $\Delta$ is a scalar to be determined self-consistently. Carrying
out a Hubbard-Stratonovich transformation we obtain the mean field
Hamiltonian 
\begin{align}
H= & \int\!\left(\rmd\boldsymbol{k}\right)\epsilon(\boldsymbol{k})\,c_{\boldsymbol{k}}^{\dagger}c_{\boldsymbol{k}}^{\,}\nonumber \\
 & +\!\!\!\sum_{\boldsymbol{k}\in\frac{1}{2}\,\mathrm{B.Z.}}\!\!\Delta\left(\boldsymbol{k}\right)c_{\boldsymbol{k}}^{\dagger}c_{-\boldsymbol{k}}^{\dagger}+\mathrm{h.c.}+\frac{2}{3g}|\Delta|^{2}.\label{eq:Mean_field_Hamiltonian}
\end{align}
The integration over the fermions (keeping $\Delta$ fixed) yields
the free energy 
\begin{equation}
F=-T\int\left(\rmd\boldsymbol{k}\right)\ln\left(\cosh\left(\frac{\beta E\left(\boldsymbol{k}\right)}{2}\right)\right)+\frac{2}{3g}\left|\Delta\right|^{2},\label{eq:Free_Energy}
\end{equation}
where $E\left(\boldsymbol{k}\right)\equiv\sqrt{\epsilon(\boldsymbol{k})^{2}+\left|\Delta\left(\boldsymbol{k}\right)\right|^{2}}$
and $T=1/\beta$ is the temperature (we set $k_{{\rm B}}=1$). The
saddle-point equation, obtained by taking the variation of $F$ with
respect to $\Delta^{*}$~\cite{key-11}, reads 
\begin{equation}
\frac{2}{3g}=\frac{1}{4}\int\left(\rmd\boldsymbol{k}\right)\tanh\left(\frac{\beta E\left(\boldsymbol{k}\right)}{2}\right)\frac{\hat{k}_{x}^{2}+\hat{k}_{y}^{2}}{E\left(\boldsymbol{k}\right)}.\label{eq:Slelf_consistency_equilibrium}
\end{equation}
Notice the extra factor of ${1}/{2}$ compared to the regular BCS
theory which stems from the fact that only one ``spin species''
is considered. To estimate the integral in Eq.~(\ref{eq:Slelf_consistency_equilibrium}),
we use the fact that close to the surface ${\cal S}_{0}$, the dispersion
relation can be Taylor-expanded as $\epsilon=\Lambda+\kappa\,k_{\perp}^{2}+\dots$,
where $k_{\perp}$ is the momentum perpendicular to ${\cal S}_{0}$
and $\kappa={1}/{2m^{*}}>0$. $\Lambda>0$, the distance of the band
edge to the chemical potential, will play a key role in what follows.
Again, notice that $\epsilon>0$, so the chemical potential never
intercepts the band.

In 3D, the self-consistency equation~(\ref{eq:Slelf_consistency_equilibrium})
becomes 
\begin{align}
\frac{2}{3g}\approx & \frac{k_{0}^{2}}{4\left(2\pi\right)^{2}}\int_{0}^{\pi}\rmd\theta\sin^{3}\theta\label{eq:simplified_selfconsistency}\\
 & \times\int\rmd k_{\perp}\frac{\tanh\left(\frac{\beta}{2}\sqrt{\left(\Lambda+\kappa k_{\perp}^{2}\right)^{2}+|\Delta|^{2}\sin^{2}\theta}\right)}{\sqrt{\left(\Lambda+\kappa k_{\perp}^{2}\right)^{2}+|\Delta|^{2}\sin^{2}\theta}}.\nonumber 
\end{align}
In a standard BCS approach, the cut-off is typically set by the Debye
frequency. Here, the divergence of the density of states at the band
edge makes this scale irrelevant and we can simply afford to extend
the $k_{\perp}$ integrals to infinity since they are convergent.


The superconducting phase transition is located at $\Delta=0$, \textit{i.e.},
\begin{align}
1 & =\frac{g\,k_{0}^{2}}{2\left(2\pi\right)^{2}}\int_{-\infty}^{\infty}\rmd k_{\perp}\,\frac{\tanh\left(\frac{\beta}{2}\left(\Lambda+\kappa k_{\perp}^{2}\right)\right)}{\Lambda+\kappa k_{\perp}^{2}}\label{eq:Transition_temperature}\\
 & \gtrsim\frac{g\,k_{0}^{2}}{2\left(2\pi\right)^{2}}\frac{\pi}{\sqrt{\Lambda\kappa}}\tanh\left(\beta\Lambda\right).\label{eq:Self_consitency}
\end{align}
The existence of a superconducting phase is therefore conditioned
by 
\begin{equation}
g>g_{\rmc}=\frac{8\pi\sqrt{\Lambda\kappa}}{k_{0}^{2}}.\label{eq:treshhold_equilibrium}
\end{equation}
Notice that the non-zero value of $k_{0}$ for this type of the band
geometry is \textit{essential} to give a finite value for the critical
coupling $g_{\rmc}$. Recall that in the case where the band geometry
derives from the spin-orbit interaction, the value of $k_{0}$ is
proportional to $|\alpha|$ as given by Eq.~(\ref{eq:k_0}). Also
notice that the threshold condition can be satisfied even with rather
pessimistic estimates: for $\Lambda\approx0.1~e$V, $\kappa\approx\left(10^{-6}~e{\mathrm{V}}\right)^{-1}c^{2}$,
and $gk_{0}^{2}\approx10^{-2}c$, the threshold condition is satisfied
with $1>0.79$. We note that in the optimistic limit where $\Lambda\rightarrow0$
we have $g_{c}\rightarrow0$ and the critical temperature is given
by $\beta_{c}=859\kappa\frac{1}{g^{2}k_{0}^{4}}$. Furthermore the
superconducting order parameter at zero temperature is given by $\Delta_{0}=\frac{g^{2}k_{0}^{4}}{\kappa}\left(\frac{3\sqrt{\pi}\Gamma^{2}\left(\frac{5}{4}\right)}{2\left(2\pi\right)^{2}}\right)^{2}$,
this leads to $\beta_{c}\Delta_{0}=2.63$.

For a more realistic limit with a sizable $\Lambda$ the transition
temperature grows as the coupling exceeds the threshold, 
\begin{equation}
T_{c}\approx\frac{\Lambda}{\tanh^{-1}({g_{c}}/{g})},\quad{\rm for}\;g>g_{\rmc}\;.
\end{equation}

The magnitude of the superconducting gap can be estimated by focusing
on the low-temperature regime $T\ll\Lambda$ for which the gap equation
Eq.~(\ref{eq:simplified_selfconsistency}) yields { 
\begin{align}
1 & \approx\frac{g}{g_{c}}\;\frac{\sqrt{2}}{5}\;\frac{\frac{9}{2}x^{2}+\sqrt{1+x^{2}}-1}{x^{2}\sqrt{\sqrt{1+x^{2}}+1}},\label{eq:Consistency_Equation}
\end{align}
with $x\equiv\Delta/\Lambda$}, from which it follows that a superconducting
gap on the order of $\Delta\sim\Lambda$ can be achieved when the
ratio $g/g_{c}$ starts to increase away from unity. Near the threshold,
the superconducting order parameter scales as 
\begin{align}
\Delta\approx\Lambda\,2\sqrt{5/3}\sqrt{1-\frac{g_{c}}{g}},\quad{\rm for}\;g>g_{\rmc}\;.\label{eq:near-threshold}
\end{align}

\paragraph{\label{sub:No-spin-orbit} Singlet case.}

We now extend the previous analysis to cases in which the Mexican
hat band geometry is not a consequence of spin-orbit coupling but
finds its origin in another mechanism. In this case, the bands are
spin-degenerate and an $s$-wave pairing, $\Delta\left(\boldsymbol{k}\right)=\Delta\left(-\boldsymbol{k}\right)$,
is allowed by symmetry. Thus, the dominant pairing channel sees a
potential $V\left(\boldsymbol{q}\right)=V\left(-\boldsymbol{q}\right)$,
and the pairing now occurs between the two spin bands.

The self-consistency equation is now given by 
\begin{equation}
\frac{1}{g}=\frac{1}{2}\int\left(\rmd\boldsymbol{k}\right)\frac{1}{E\left(\boldsymbol{k}\right)}\tanh\left(\frac{\beta E\left(\boldsymbol{k}\right)}{2}\right),\label{eq:Self_Consistency relation}
\end{equation}
where $E(\boldsymbol{k})\equiv\sqrt{\epsilon(\boldsymbol{k})^{2}+|\Delta|^{2}}$.
The phase transition ($\Delta=0$) occurs at 
\begin{align}
1 & =\frac{g\,k_{0}^{2}}{\left(2\pi\right)^{2}}\int_{-\infty}^{\infty}\rmd k_{\perp}\frac{\tanh\left(\frac{\beta}{2}\left(\Lambda+\kappa k_{\perp}^{2}\right)\right)}{\Lambda+\kappa k_{\perp}^{2}}\label{eq:Transition_temperature-1}\\
 & \gtrsim\frac{g\,k_{0}^{2}}{\left(2\pi\right)^{2}}\frac{\pi}{\sqrt{\Lambda\kappa}}\tanh\left(\beta\Lambda\right).\label{eq:self_consistency}
\end{align}
The existence of a superconducting phase is now conditioned by 
\begin{equation}
g>g_{\rmc}=\frac{4\pi\sqrt{\Lambda\kappa}}{k_{0}^{2}}.\label{eq:treshhold_equilibrium-1}
\end{equation}
The analysis of this case is analogous to the previous: once $g$
exceeds the threshold $g_{c}$, a superconducting phase appears. The
critical temperature can be of order $\Lambda$ once the threshold
starts to be exceeded. We note that in the optimistic limit where
$\Lambda\rightarrow0$ we have $g_{c}\rightarrow0$ and the critical
temperature is given by $\beta_{c}=215\kappa\frac{1}{g^{2}k_{0}^{4}}$.
Furthermore the superconducting order parameter at zero temperature
is given by $\Delta_{0}=\frac{g^{2}k_{0}^{4}}{\kappa}\left(\frac{8\Gamma^{2}\left(\frac{5}{4}\right)}{\left(2\pi\right)^{2}\sqrt{\pi}}\right)^{2}$,
this leads to $\beta_{c}\Delta_{0}=1.90$.

For the more realistic limit where $\Lambda$ is sizable the magnitude
of the gap can be estimated again by focusing on the low-temperature
regime when $T\ll\Lambda$ and using the gap equation, yielding {
\begin{equation}
1=\frac{g}{g_{c}}\;\frac{2}{\pi}\,\mathrm{E}_{K}\left(\frac{1}{\sqrt{2}}\sqrt{1-1/\sqrt{1+x^{2}}}\right)\big{/}\left(1+x^{2}\right)^{1/4},\label{eq:gap_equation-2}
\end{equation}
with $x\equiv|\Delta|/\Lambda$ and $\mathrm{E}_{K}$ is the complete
elliptic integral of the first kind.} Again, it follows that a superconducting
gap on the order of $\Delta\sim\Lambda$ can be achieved when the
ratio $g/g_{c}$ starts to increase away from unity, and near the
threshold we obtain 
\begin{align}
\Delta\approx\Lambda\,\frac{4}{\sqrt{3}}\sqrt{1-\frac{g_{c}}{g}},\quad{\rm for}\;g>g_{\rmc}\;.\label{eq:near-threshold-s-wave}
\end{align}

\label{sec:Critical-temperature}\textit{ Strong-coupling approach.}
Let us now give a description beyond weak-coupling mean-field theory
by means of a Migdal-Eliashberg approach~\cite{Eliashberg}. This
consists in including explicitly the screened electron-phonon interaction
and the screened Coulomb interaction, and establishing a self-consistent
equation on the resulting self-energy, yielding an estimate of the
critical temperature. If the phonon frequency is much smaller than
the electronic energy scale, Migdal's theorem states that the phononic
vertex corrections can be neglected, even if the electron-phonon coupling
constant is large~\cite{Migdal,Jong,key-12}. We derive a version
of this theorem applicable to our case in~\cite{supplementary}.

To simplify, we consider the second scenario described above, \textit{i.e.},
the $s$-wave pairing case in the absence of spin-orbit coupling.
The $p$-wave case with spin-orbit coupling is conceptually identical
and, in the strong-coupling limit, can be shown to have the same critical
temperature up to numerical factors of order unity. However, as the
precise form of the spin-orbit coupling is not available, these factors
cannot be reliably computed.

Let us start with the following Hamiltonian 
\begin{align}
H= & \int\!\left(\rmd\boldsymbol{k}\right)\left(\epsilon(\boldsymbol{k})c_{\boldsymbol{k}}^{\dagger}c_{\boldsymbol{k}}^{\,}+V_{{\rm C}}\left(\boldsymbol{k}\right)\rho_{\boldsymbol{k}\downarrow}^{\dagger}\rho_{\boldsymbol{k}\uparrow}+\Omega_{\boldsymbol{k}}b_{\boldsymbol{k}}^{\dagger}b_{\boldsymbol{k}}^{\,}\right)\nonumber \\
 & +\sum_{\sigma}\int\!\left(\rmd\boldsymbol{k}\right)\left(\rmd\boldsymbol{q}\right)g_{\boldsymbol{q}}\;c_{\boldsymbol{k}+\boldsymbol{q},\sigma}^{\dagger}c_{\boldsymbol{k},\sigma}^{\,}\left(b_{\boldsymbol{q}}^{\,}+b_{-\boldsymbol{q}}^{\dagger}\right).\label{eq:Main_Hamiltonian-1}
\end{align}
Here, $\rho_{\boldsymbol{k}\uparrow}=\int\left(\rmd\boldsymbol{q}\right)c_{\boldsymbol{q}\uparrow}^{\dagger}c_{\boldsymbol{q}+\boldsymbol{k}\uparrow}^{\,}$
is the electron density, $V_{{\rm C}}\left(\boldsymbol{k}\right)$
is the screened Coulomb interaction, and $g_{\boldsymbol{q}}$ is
the electron phonon-coupling matrix (which is also screened). Let
us introduce the phonon propagator in Matsubara time~\cite{key-12}
\begin{align}
 & D\left(\boldsymbol{q},\omega_{n}\right)=\nonumber \\
 & -\int_{0}^{\beta}\!\!\!\rmd\tau\,\rme^{\rmi\omega_{n}\tau}\left\langle T_{\tau}\left(b_{\boldsymbol{q}}(\tau)+b_{-\boldsymbol{q}}^{\dagger}(\tau)\right)\left(b_{-\boldsymbol{q}}(0)+b_{\boldsymbol{q}}^{\dagger}(0)\right)\right\rangle \nonumber \\
 & =-\frac{2\Omega_{\boldsymbol{q}}}{\omega_{n}^{2}+\Omega_{\boldsymbol{q}}^{2}}\approx-\frac{2}{\Omega_{\boldsymbol{q}}}\delta_{n,0}.\label{eq:Propagator}
\end{align}
Here, $\omega_{n}=n\,{2\pi}/\beta$ and in the last equality we have
taken the high temperature limit $T\gg\Omega_{q}$.

Below, we derive a self-consistency equation for the pairing amplitude.
It is convenient to introduce the standard Nambu Green's functions~\cite{key-13}:
\begin{align}
 & G\left(\boldsymbol{k},\tau\right)=\label{eq:Nambu_Function}\\
 & -\left(\begin{array}{cc}
\left\langle T_{\tau}c_{\boldsymbol{k}\uparrow}^{\,}(\tau)c_{\boldsymbol{k}\uparrow}^{\dagger}(0)\right\rangle  & \left\langle T_{\tau}c_{\boldsymbol{k}\uparrow}^{\,}(\tau)c_{-\boldsymbol{k}\downarrow}^{\,}(0)\right\rangle \\
\left\langle T_{\tau}c_{-\boldsymbol{k}\downarrow}^{\dagger}(\tau)c_{\boldsymbol{k}\uparrow}^{\dagger}(0)\right\rangle  & \left\langle T_{\tau}c_{-\boldsymbol{k}\downarrow}^{\dagger}(\tau)c_{-\boldsymbol{k}\downarrow}^{\,}(0)\right\rangle 
\end{array}\right).\nonumber 
\end{align}
The corresponding self-energy obeys the matrix equation 
\begin{equation}
\Sigma\left(\boldsymbol{k},\omega_{n}\right)=G_{0}^{-1}\left(\boldsymbol{k},\omega_{n}\right)-G^{-1}\left(\boldsymbol{k},\omega_{n}\right),\label{eq:Self_Energy}
\end{equation}
with the non-interacting Green's function $G_{0}\left(\boldsymbol{k},\omega_{n}\right)=\left[\rmi\omega_{n}\tau_{0}-\epsilon(\boldsymbol{k})\tau_{3}\right]^{-1}$.
$\tau_{0}$ and $\tau_{1}$, $\tau_{2}$, $\tau_{3}$ denote the identity
and Pauli matrices in the Nambu space. To leading order, the self-energy
is given by 
\begin{align}
\Sigma\left(\boldsymbol{k},\omega_{n}\right)= & -T\sum_{\omega_{n'}}\int\!\left(\rmd\boldsymbol{q}\right)\tau_{3}G\left(\boldsymbol{k}-\boldsymbol{q},\omega_{n'}\right)\tau_{3}\nonumber \\
 & \times\left(|g_{\boldsymbol{q}}|^{2}D\left(\boldsymbol{q},\omega_{n}-\omega_{n'}\right)+V_{{\rm C}}\left(\boldsymbol{q}\right)\right).\label{eq:Green's_function}
\end{align}
Next, we note that as long as our system is isotropic, \textit{e.g.}
$\epsilon(\boldsymbol{k})=\epsilon(k)$, $g_{\boldsymbol{q}}=g_{q}$,
$\Omega_{\boldsymbol{q}}=\Omega_{q}$, $V_{{\rm C}}\left(\boldsymbol{q}\right)=V_{{\rm C}}\left(q\right)$,
and the quantities $V_{{\rm C}}\left(q\right)$ and ${|g_{q}|^{2}}/{\Omega_{q}}$
do not have strong dependence on $q$ for $q\sim\sqrt{{\Lambda}/{\kappa}}$
then the self-energy does not depend on the wavevector, \textit{e.g.}
$\Sigma\left(k,\omega_{n}\right)\approx\Sigma\left(\omega_{n}\right)$.
On symmetry grounds, the self-energy can be decomposed as~\cite{key-13}
\begin{equation}
\Sigma\left(\omega_{n}\right)=\left(1-Z\right)\rmi\omega_{n}\tau_{0}+\Delta(\omega_{n})\tau_{1}+\chi\tau_{3}.\label{eq:self_energy}
\end{equation}
Using Eq.~(\ref{eq:Self_Energy}), we obtain 
\begin{equation}
G\left(k,\omega_{n}\right)=-\frac{Z\rmi\omega_{n}{\tau_{0}}+\Delta(\omega_{n})\tau_{1}+\left(\epsilon\left(k\right)+\chi\right)\tau_{3}}{Z^{2}\omega_{n}^{2}+\left(\epsilon\left(k\right)+\chi\right)^{2}+\Delta(\omega_{n})^{2}}.\label{eq:Green's_function-1}
\end{equation}
For simplicity, we assume that $Z=1$ and $\chi=0$ (the former renormalizes
$\Lambda$ and can always be set to zero). Using Eq.~(\ref{eq:Green's_function})
in Eq.~(\ref{eq:self_energy}), we obtain the self-consistency equation
\begin{align}
\Delta\left(\omega_{n}\right) & =\lambda T\int\rmd k\frac{\Delta\left(\omega_{n}\right)}{\omega_{n}^{2}+\epsilon(k)^{2}+\Delta(\omega_{n})^{2}}-\nonumber \\
 & -\mu_{{\rm C}}T\sum_{\omega_{n'}}\int\rmd k\frac{\Delta\left(\omega_{n'}\right)}{\omega_{n'}^{2}+\epsilon(k)^{2}+\Delta(\omega_{n'})^{2}},\label{eq:Self_consistency}
\end{align}
where we introduced 
\begin{align}
\mu_{{\rm C}}\equiv\frac{1}{\left(2\pi\right)^{3}}\int_{{\cal S}_{0}}\!\!\!V_{{\rm C}}\left(\boldsymbol{k}\right)\mbox{ and }\lambda\equiv\frac{1}{\left(2\pi\right)^{3}}\int_{{\cal S}_{0}}\!\!\!\left|g_{\boldsymbol{q}}\right|^{2}\frac{2}{\Omega_{\boldsymbol{q}}},\label{eq:Averaged_values}
\end{align}
respectively the Coulomb and phonon interactions integrated over the
surface ${\cal S}_{0}$. Close to the critical temperature, we may
drop the $\Delta\left(\omega_{n}\right)^{2}$ terms in the denominators
of in Eq.~(\ref{eq:Self_consistency}) which, after integration over
$k$ using $\epsilon(k)=\Lambda+\kappa k^{2}+\ldots$, can be recast
as 
\begin{equation}
\Delta\left(\omega_{n}\right)=A_{n}\Delta\left(\omega_{n}\right)-\frac{\mu_{{\rm C}}}{\lambda}\sum_{n'}A_{n'}\Delta\left(\omega_{n'}\right),\label{eq:Self_Consistency_simplified}
\end{equation}
with $A_{n}\equiv\pi\frac{\lambda T}{\sqrt{\kappa}}[\sqrt{\Lambda^{2}+\omega_{n}^{2}}(\sqrt{\Lambda+\rmi\omega_{n}}+\sqrt{\Lambda-\rmi\omega_{n}})]^{-1}$.
\begin{equation}
1=\frac{\mu_{{\rm C}}}{\lambda}\sum_{n}\frac{A_{n}}{A_{n}-1}.\label{eq:Self_consistency_final}
\end{equation}
This equation admits a non-trivial solution whenever $\lambda\geq\lambda_{\rmc}\left({\mu_{{\rm C}}}/{\lambda}\right)$,
with $\lambda_{\rmc}\left(0\right)<\lambda_{\rmc}\left({\mu_{{\rm C}}}/{\lambda}\right)<\lambda_{\rmc}\left(\infty\right)$.
Computing explicitly $\lambda_{\rmc}\left(0\right)=2^{3/2}\sqrt{\kappa\Lambda}$
and $\lambda_{\rmc}\left(\infty\right)=3\cdot2^{3/2}\sqrt{\kappa\Lambda}$,
this implies that the critical temperature always satisfies $T_{\rmc}>{\Lambda}/{\sqrt{3\pi^{2}}}$.

\paragraph{\label{sec:Conclusions}Conclusions.}

We have studied the emergence of superconductivity for those semiconductors
with a band in the shape of a Mexican hat, where the energy reaches
an extremum on a band-edge surface ${\cal S}_{0}$. We have set the
chemical potential for the semiconductor close to the band edge, but
not intercepting the band. Therefore, in the absence of interactions,
the system would have vanishing conductivity at zero temperature.
We have shown, both through a mean field and through a strong coupling
calculation, that phonon-mediated superconductivity arises and is
robust to high temperatures. The mechanism benefits from a quasi one-dimensional
divergent density of states at the band edge, making an ``economical''
use of the energy levels on a shell near the band extrema.

This work has been supported by the Rutgers CMT fellowship (G.G.),
the NSF grant DMR-115181 (C.A.), and the DOE Grant DEF-06ER46316 (C.C.).

\widetext
\newpage
\begin{center}
\textbf{\large Supplemental material to ``Band-edge superconductivity''}
\end{center}
\setcounter{equation}{0}
\setcounter{figure}{0}
\setcounter{table}{0}
\setcounter{page}{1}
\makeatletter
\renewcommand{\theequation}{S\arabic{equation}}
\renewcommand{\thefigure}{S\arabic{figure}}
\renewcommand{\bibnumfmt}[1]{[S#1]}
\renewcommand{\citenumfont}[1]{S#1}

\begin{figure}[h]
\begin{centering}
\includegraphics[scale=0.3]{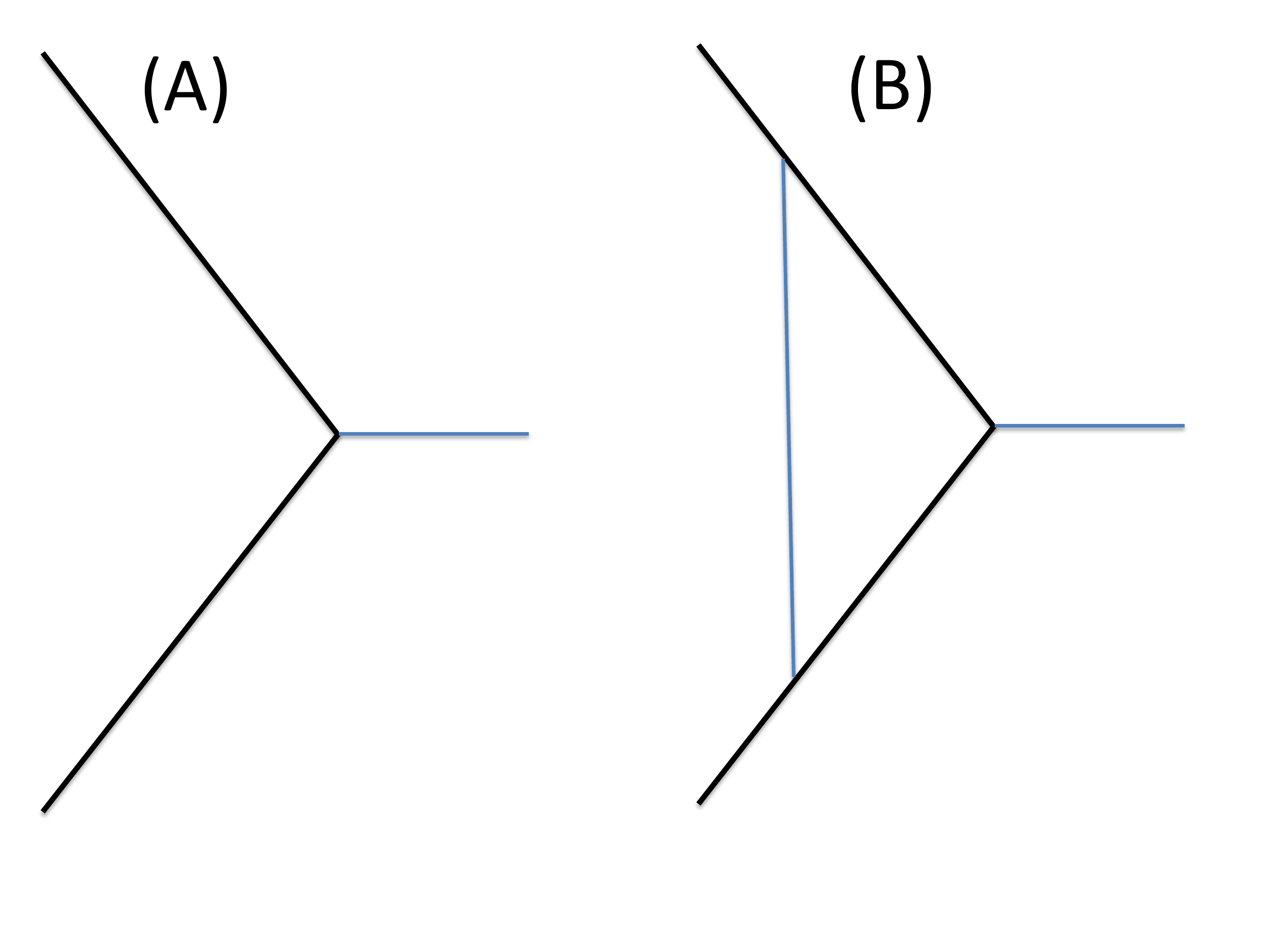} 
\par\end{centering}

\protect\protect\protect\protect\protect\protect\protect\caption{\label{fig:Migdal} The contribution of diagram (A) to the self energy
is much larger than the one of diagram (B).}
\end{figure}

\subsection*{Migdal's theorem}

The above derivation relied on the fact that only a certain class
of diagrams, the ones with just one electron line, give the dominant
contribution to the electron self energy. In this note, we show that
this analogue to Migdal's theorem holds for the band structure considered
in this paper. In particular, ignoring numerical factors on the order
of unity, we show that the ratio of the diagram shown in Fig.~\ref{fig:Migdal}(a)
to the one given in Fig.~\ref{fig:Migdal}(b) is given by: 
\begin{equation}
R_{a/b}\sim k_{0}\sqrt{\frac{\kappa}{\Lambda}}\gg1\;.\label{eq:Ratio}
\end{equation}

Indeed, it rather simple to see that the value of the diagram given
in Fig.~\ref{fig:Migdal}(a) is given by $g_{q}$ while the value
of the diagram in Fig.~\ref{fig:Migdal}(b) is proportional to $k_{0}\frac{\left|g_{k}\right|^{2}}{\Omega_{k}}\frac{T}{\Lambda^{2}+T^{2}}\frac{\Lambda}{\kappa}g_{q}$.
Here, $q$ is the incoming momentum and $k$ is an arbitrary momentum.
Moreover, according to the-self consistency equation {[}see Eq.~(29)
in the main text{]} we have that: $1\sim k_{0}^{2}\frac{\left|g_{k}\right|^{2}}{\Omega_{k}}\frac{T}{\Lambda^{2}+T^{2}}\sqrt{\frac{\Lambda}{\kappa}}$.
Therefore, Eq.~(\ref{eq:Ratio}) follows automatically. 
\end{document}